\documentclass[final,3p,times,twocolumn]{elsarticle}

\usepackage{lineno,hyperref}
\modulolinenumbers[5]

\usepackage[T1]{fontenc}
\usepackage{graphicx}
\usepackage{epstopdf}
\usepackage{color,soul}
\journal{.}

\begin{document}

\begin{frontmatter}

\title{Testing the capability of low-energy light ions identification \\of the TRACE silicon detectors}


\author[a,b]{N. Cieplicka-Ory\'nczak\corref{mycorrespondingauthor}}
\cortext[mycorrespondingauthor]{Corresponding author}
\ead{natalia.cieplicka@ifj.edu.pl}

\author[d,e]{D. Mengoni}
\author[a]{M. Ciema\l{}a}
\author[b,c]{S. Leoni}
\author[a]{B. Fornal}
\author[f]{J.A. Due\~nas}
\author[b,c]{S. Brambilla}
\author[b,c]{C. Boiano}
\author[d,e]{P.R. John}
\author[d,e]{D. Bazzacco}
\author[b]{G. Benzoni}
\author[b,c]{G. Bocchi}
\author[b,c]{S. Capra}
\author[b,c]{F.C.L. Crespi}
\author[g]{A. Goasduff}
\author[g]{K. Hady\'nska-Kl\k{e}k}
\author[a]{\L{}.W. Iskra}
\author[g]{G. Jaworski}
\author[d,e]{F. Recchia}
\author[d,e]{M. Siciliano}
\author[d,e]{D. Testov}
\author[g]{J.J. Valiente-Dob\'on}

\address[a]{Institute of Nuclear Physics Polish Academy of Sciences, PL-31342 Krakow, Poland}
\address[b]{Istituto Nazionale di Fisica Nucleare Sezione di Milano, Via Celoria 16, I-20133 Milano, Italy}
\address[c]{Universit\`a degli Studi di Milano, Via Celoria 16, 20133 Milano, Italy}
\address[d]{Istituto Nazionale di Fisica Nucleare Sezione di Padova, I-35131 Padova, Italy}
\address[e]{Dipartimento di Fisica dell'Universita degli Studi di Padova, I-35131 Padova, Italy}
\address[f]{Departmento de Ingenier\'ia El\'ectrica, ETSI Universidad de Huelva, $21819$ Huelva, Spain}
\address[g]{Istituto Nazionale di Fisica Nucleare Laboratori Nazionali di Legnaro, I-35020 Legnaro, Italy}

%---------------------------------------------------------------------------------------------
%-----------------------------------   ABSTRACT   --------------------------------------------
%---------------------------------------------------------------------------------------------
\begin{abstract}
The in-beam tests of two Si pixel type TRACE detectors have been performed at Laboratori Nazionali di Legnaro (Italy). The aim was to investigate the possibility of identifying heavy-ion reactions products with mass $A\sim10$ at low kinetic energy, i.e., around 10~MeV. Two separate read-out chains, digital and analog, were used. The Pulse Shape Analysis technique was employed to obtain the identification matrices for the digitally processed part of the data. Separation in both charge and mass was obtained, however, the $\alpha$ particles contaminated significantly the recorded data in the lower energy part. Due to this effect, the identification of the light products ($^{7,6}$Li isotopes) could be possible down only to $\sim$20~MeV 
\end{abstract}
%---------------------------------------------------------------------------------------------


\end{frontmatter}

%\linenumbers

%---------------------------------------------------------------------------------------------
%-------------------------------------   INTRO   ---------------------------------------------
%---------------------------------------------------------------------------------------------
\section{Introduction} 

One of the modern detection methods, offering identification of the reaction products with very low energy, is the Pulse Shape Analysis (PSA). This method can be applied to the signals from silicon detectors. As shown by Mengoni~\textit{et al.}~\cite{ref4}, in the case of the TRACE array~\cite{ref4}, consisting of 200-$\mu$m thick silicon modules and divided in 60 separately read pixels, the identification of the $^{1,2,3}$H isotopes can be easily obtained. In addition, the separation between $^3$He and $^4$He was also observed. This proved that the thin, floating-zone technique (FZ) detector, with the uniformity guaranteed by the fine pad segmentation, may provide a good particle discrimination when the PSA technique is applied.

In this paper, we present the results of an experiment performed at the Legnaro National Laboratory which has as the objective an in-beam test of the possibility of identification of reaction products with mass A$\sim$10 at low kinetic energy (around 10~MeV) by Pulse Shape Analysis technique with the segmented silicon detectors of the TRACE setup. If successful, this would allow to investigate the structure of light reaction products, such as Be, B, C, N, O, by a direct measurement of their energy, position, mass, and
charge.

%---------------------------------------------------------------------------------------------
%-------------------------------------   SET UP   --------------------------------------------
%---------------------------------------------------------------------------------------------
\section{Experimental set up and procedures}
The experiment was performed in July 2016 at the Legnaro National Laboratory (Italy). A $^{37}$Cl beam, with energy of 186~MeV and intensity around 1~pnA was focused on a 0.1~mg/cm$^2$ thick $^{12}$C target. Two TRACE silicon detectors were placed inside the GALILEO HPGe array scattering chamber. Each of them made up of 60, 4$\times$4~mm$^2$ pads with 4.5~mm pitch junction side, forming a matrix of 12$\times$5 pads covering an area of approximately 50$\times$20 mm$^2$. A common electrode covered the entire ohmic side of the detector. The thickness of the detectors was 200~$\mu$m. However, due to the limited number of available read-out channels, only the signals from the common electrode and a few pads or groups of pads were recorded. The grouping of pixels was done on the PCB in the same way for both detectors: a group of 8, two groups of 4, two groups of 2 pads and three single pads were used per detector.

Both TRACE detectors were positioned at the forward angles, with the active area covering $\sim35-57^{\circ}$, symmetrically with respect to the beam axis (where the maximum of the production of light ions was reached) and the ohmic side (the common electrode, denoted also as ``BACK'') facing the incoming particles. 

Inside the chamber, the detectors were connected with a short cable to a 16-channel charge-sensitive preamplifier, designed at INFN Milano~\cite{ref5}. The preamplifier gain was 45~mV/MeV yeliding a dynamic range of approximately 100~MeV. The trigger of the acquisition was the signal from the common electrode. The signals from one detector were sent to the digital acquisition, while the second detector was connected to the analog read-out line.

\begin{figure} 
		\centering
    \includegraphics[width=0.45\textwidth]{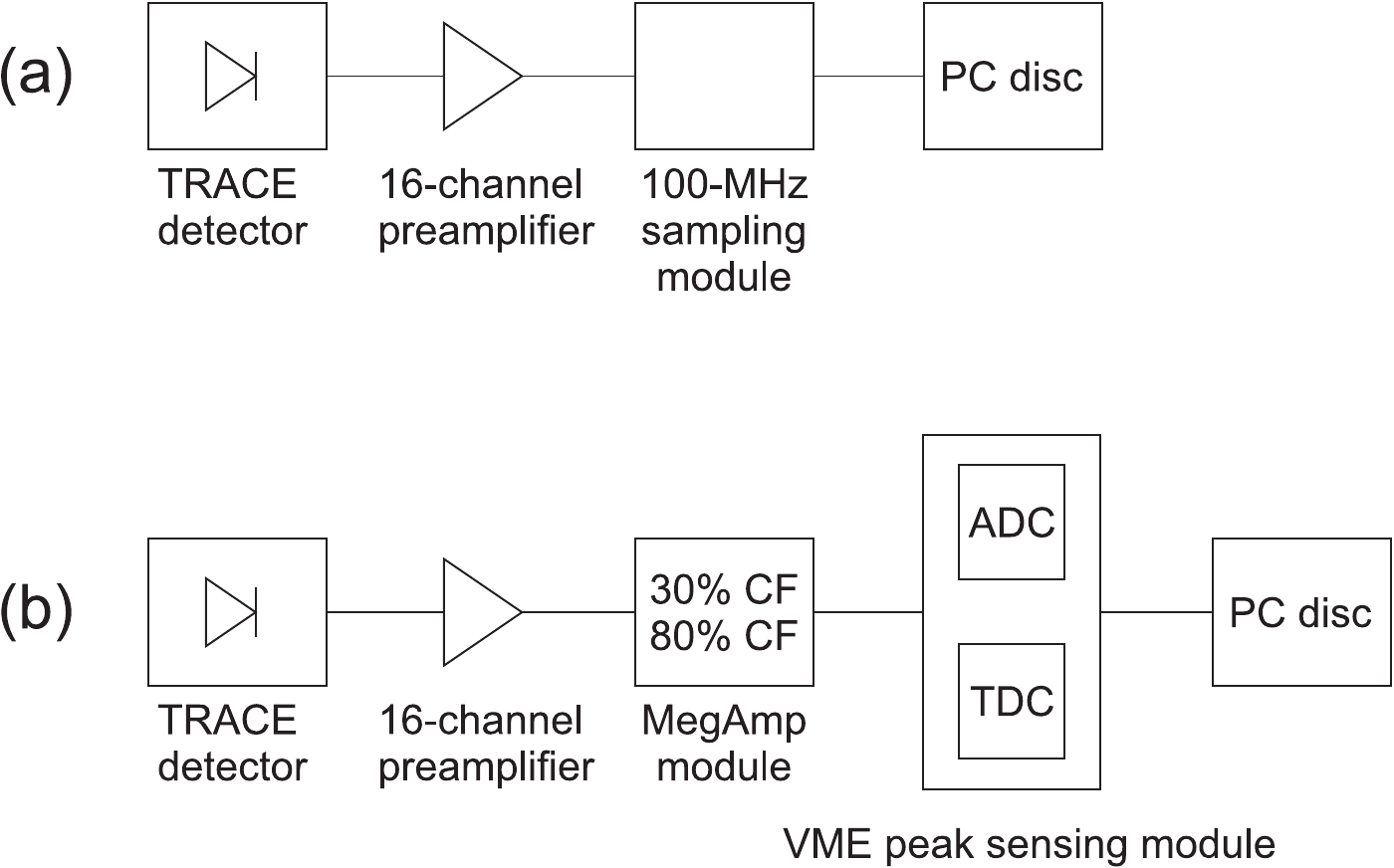}
    \caption{\label{setup} The schemes of the acquisition lines for two TRACE detectors wich were used during the experiment: (a) digital electronics, (b) analog electronics.}
\end{figure}

The chain of the digital acquisition (Fig.~\ref{setup}(a)) allowed to collect the signals (``traces'') digitized by the 100-MHz sampling module. The length of the recorded trace was 1~$\mu$s, giving 100 points, at every 10~ns each. This length of the traces was chosen to assure recording the full rise time on the one hand and at the same time minimize the amount of written data. The signals were recorded, digitized and afterwards processed to extract the relevant observables. The placement of the TRACE module in the chamber with the ohmic side facing the reaction products was important to enhance the PSA capability. The signals were collected both from the ohmic side of the detector as well as from a few groups of pads (or single pads).

The trigger of the TRACE acquisition, obtained in this case from a digital leading edge discriminator embedded in the module, was the signal from the common electrode (``BACK''). The signals from the pads were collected only if the BACK signal was present.

The preamplifier signals from the second detector were processed by an analog chain (Fig.~\ref{setup}(b)), using the
MegAmp amplifier module (also developed by INFN Milano~\cite{MegAmp}), which provided energy and two time information signals from 30\% and 80\% Constant Fraction Discriminator (CFD). Next, the signals were digitized by VME peak sensing ADC and TDC modules. The trigger for the analog acquisition was constructed as the OR of the outputs of 80\% CFD. 

%------------------------------------   FIG. 1   ---------------------------------------------
%---------------------------------------------------------------------------------------------

%---------------------------------------------------------------------------------------------
%-----------------------------------   ANALYSIS   --------------------------------------------
%---------------------------------------------------------------------------------------------
\section{Data analysis}
\subsection{Digital acquisition -- Pulse Shape Analysis}
The analog charge signal from the output of the preamplifier was digitized at 100~MHz and processed offline. The baseline was calculated from 20 points (200~ns) and subtracted from the 1-$\mu$s traces, after checking if no pile-ups were present. In order to obtain the quantities allowing to separate detected reaction products, i.e., the maximum of the current signal  ($I_{max}$) and the rise time ($T_r$), the ROOT class TSpline3 interpolation algorithm was applied, in order to ensure the first and second derivative continuity. The precision of the calculation (10~ps) was chosen as a compromise between the computing time and sufficient precision. From the interpolated function, the rise time was obtained as 30-70\% of the charge signal maximum which was calculated as the average of the 5 greatest values. The current signal was analytically extracted from the first derivative of the charge interpolated function. The 100 interpolated points were enough to calculate the stable maximum ($I_{max}$).

%------------------------------------   FIG. 2   ---------------------------------------------
\begin{figure} [h]
		\centering
    \includegraphics[width=0.45\textwidth]{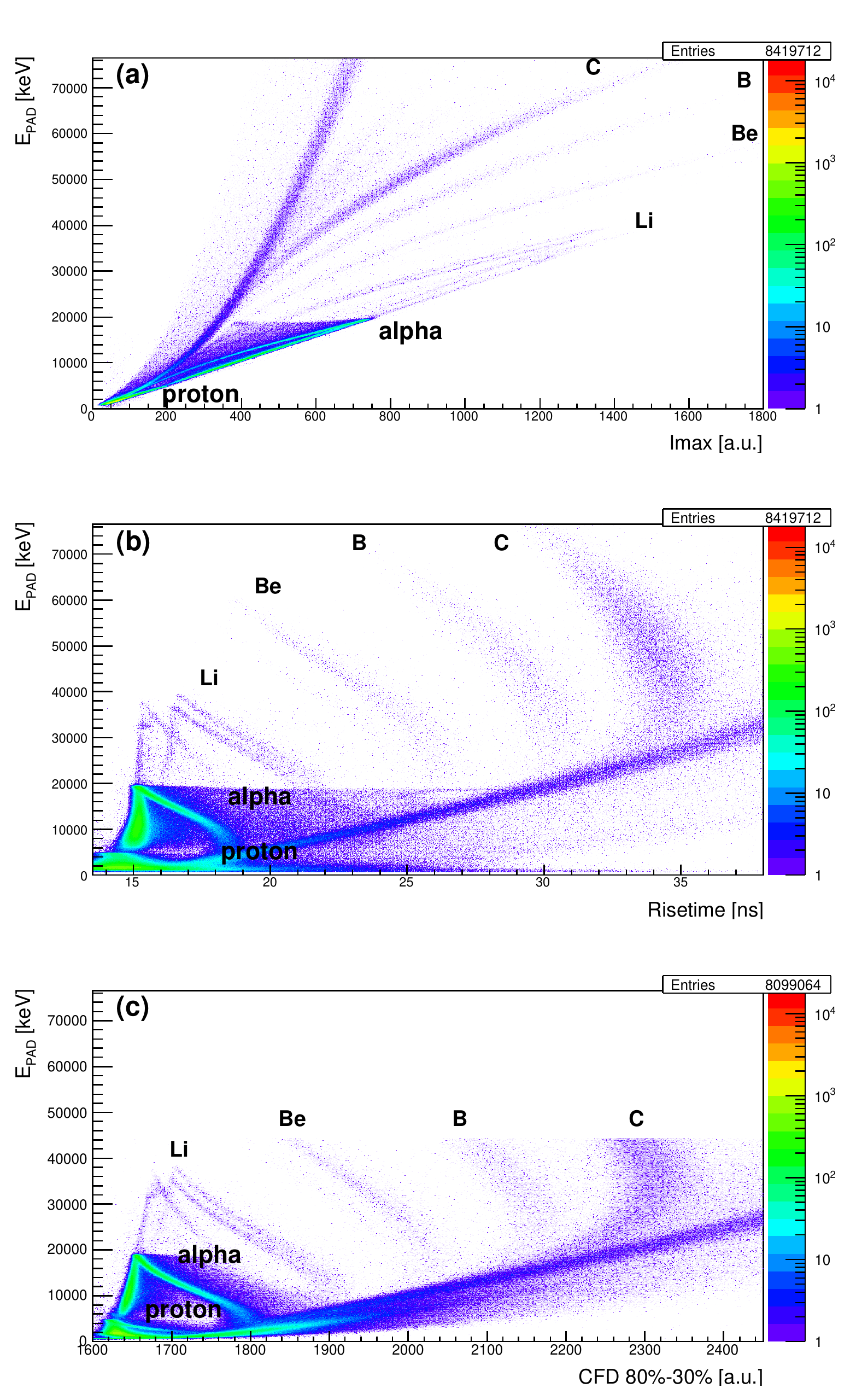}
    \caption{\label{matrices} Identification matrices constructed from the signal from a group of 2 PADs: (a) $E_{PAD}$ \textit{vs} $I_{max}$ (digital electronics), (b) $E_{PAD}$ \textit{vs} $T_r$ (digital electronics), (c) $E_{PAD}$ \textit{vs} 80-30\% CFD (analog electronics).}
\end{figure}
%---------------------------------------------------------------------------------------------

The information on the energy was obtained from the online measurements. Signals with time window of 2~$\mu$s were processed in order to find the maximum proportional to the energy of the particle. The energy resolution (30~keV at 5.5~MeV) was checked using a three-peak alpha source ($^{239}$Pu-$^{241}$Am-$^{244}$Cm). 

As a first step towards the construction of the identification matrices, the energy correlation between the ``BACK'' (ohmic) and ``PAD'' (junction) sides was done. Only the events that generated the same energy on both sides of the detector ($E_{BACK}=E_{PAD}$), i.e., the events lying on the diagonal of the $E_{BACK}$--$E_{PAD}$ correlation matrix, were taken for further analysis. Fullfilling this condition assured that only the events when one pad (or group of pads) fired were considered.  

The values of $I_{max}$ and $T_r$, derived from the interpolated waveforms and correlated with the energy, allowed to construct the identification matrices and separate the light products (Fig.~\ref{matrices}(a) and (b)).

\subsection{Analog analysis}

The information from analog acquisition of the second TRACE module provided the correlation between the energy of the particle and the time difference between the 80\% and 30\% of CFD (Fig.~\ref{matrices}(c)). Before constructing the identification matrix, the condition of $E_{BACK}=E_{PAD}$ was applied, as previously. The energy resolution was obtained using three-peak alpha source as 31~keV at 5.5~MeV.

%---------------------------------------------------------------------------------------------
%------------------------------------   RESULTS   --------------------------------------------
%---------------------------------------------------------------------------------------------
\section{Results}
The 186-MeV beam of $^{37}$Cl focused on the $^{12}$C target opened various direct and compound-nucleus reaction channels. In particular, light species, i.e., Li, Be, B isotopes, resulting from transfer reactions of a few nucleons, were produced at $\sim40-60^{\circ}$, where the detectors were placed. On the contrary, the target-like products could not be distinguished because they were out of the ADC dynamic.  

Although the best result in terms of the particle identification were expected to be obtained for electronic channels where only a single pad was connected, due to limited statistics the analysis was performed considering two connected pads positioned at average $\Theta$ angle 37$^{\circ}$ with respect to the beam. Representative examples of identification matrices presented in Fig.~\ref{matrices}(a), (b), and (c), correspond to 12~hours of measurement by the two connected pads. 

%------------------------------------   FIG. 3   ---------------------------------------------
\begin{figure} [h]
		\centering
    \includegraphics[width=0.45\textwidth]{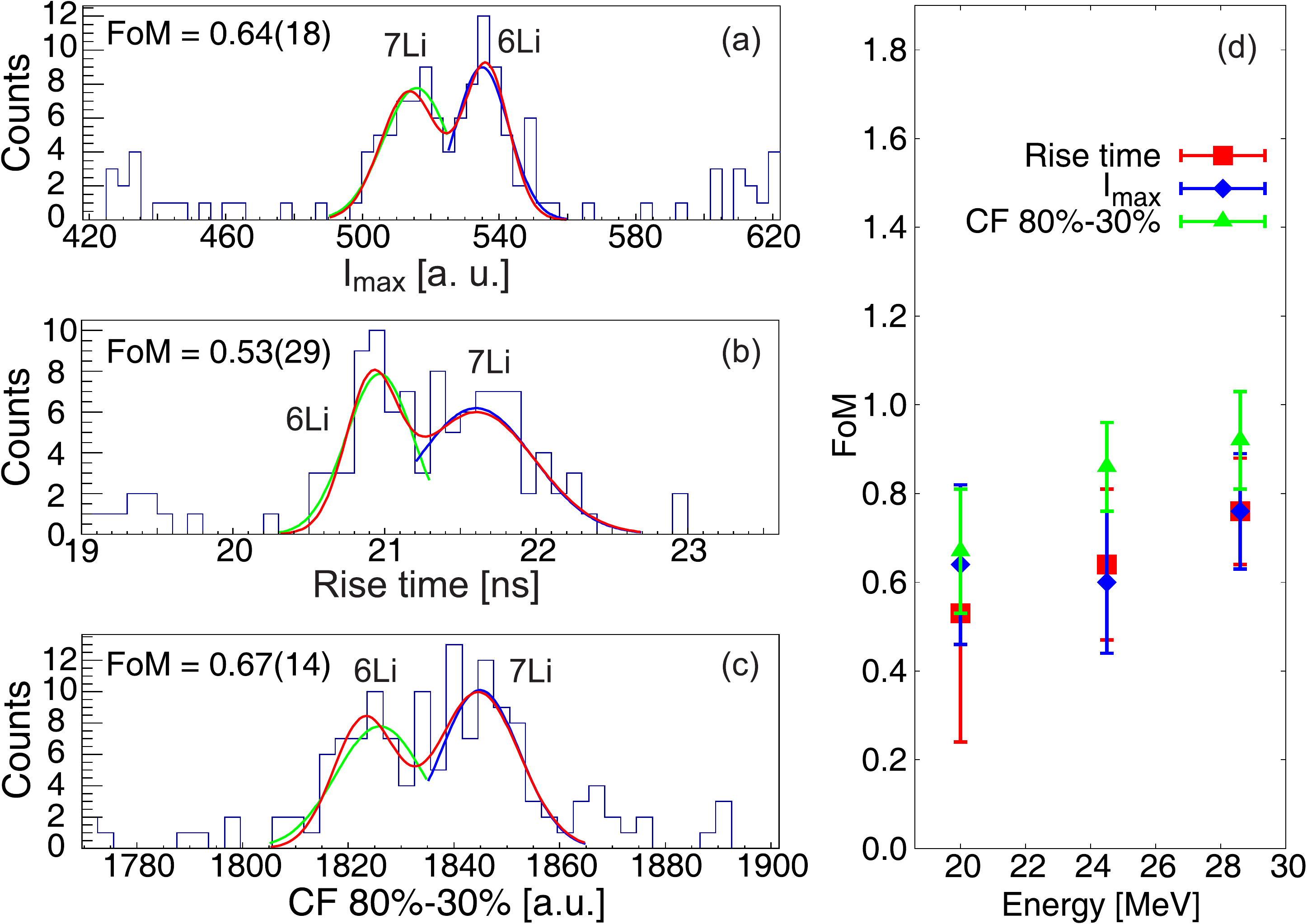}
    \caption{\label{Fig2} Identification of $^{7,6}$Li isotopes -- spectra for calculating the $FoM$ obtained by projecting the 400-keV wide cuts centered at 20~MeV from the correlation matrices for 2 PADs: (a) $E_{PAD}$ \textit{vs} $I_{max}$, (b) $E_{PAD}$ \textit{vs} $T_r$, (c) $E_{PAD}$ \textit{vs} 80-30\% CFD. The values of $FoM$ for the three identification methods calculated for the energy ranges of 19.8-20.2, 24.3-24.7, and 28.4-28.8~MeV are presented in panel (d).}
\end{figure}
%------------------------------------   TAB. 1   ---------------------------------------------
\begin{table*}
\centering
\caption{\label{tab1} The obtained with three different methods values of $FoM$ for the $^{7,6}$Li isotopes are presented (columns 2-4) for the 2-MeV wide energy cuts centered at the energy values given in column 1.}
\begin{tabular}{llcc}
\hline
\hline
$E_{PAD}$ [MeV] & $FoM$ $I_{max}$ [a.u.] & $FoM$ $T_r$ [ns] & $FoM$ 80-30\% CFD [a.u.]\\
\hline									
20.0	& 0.64(18)	& 0.53(29)	& 0.84(11)\\
\hline									
24.5	& 0.60(16)	& 0.64(17)	& 0.86(10)\\
\hline									
28.6	& 0.76(13)	& 0.76(12)	& 0.92(11)\\
\hline
\hline
\end{tabular}
\end{table*}
%------------------------------------   FIG. 4   ---------------------------------------------
\begin{figure} 
		\centering
    \includegraphics[width=0.45\textwidth]{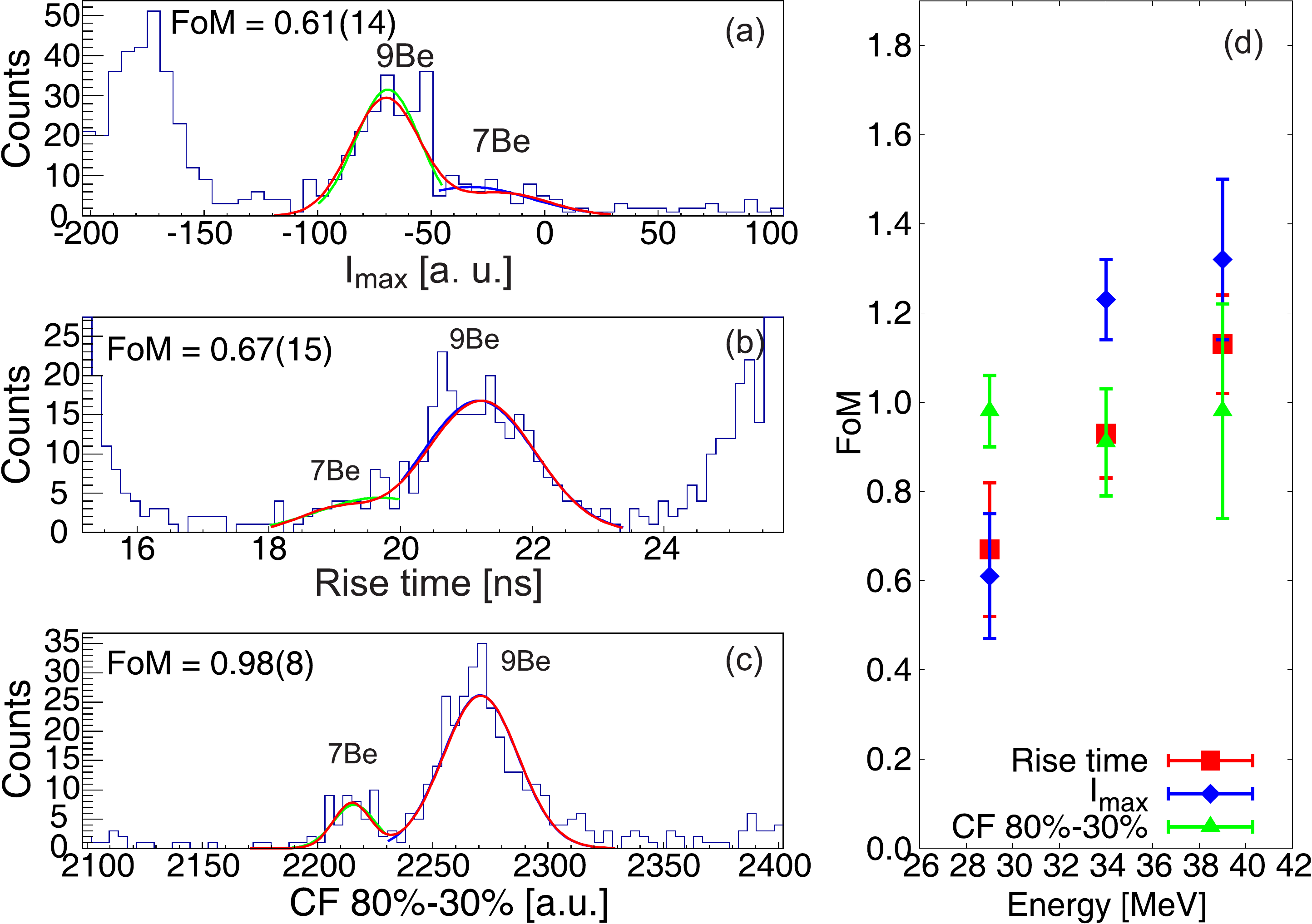}
    \caption{\label{Fig3} Identification of $^{9,7}$Be isotopes -- spectra for calculating the $FoM$ obtained by projecting the 2-MeV wide cuts centered at 29~MeV from the correlation matrices for 2 PADs: (a) $E_{PAD}$ \textit{vs} $I_{max}$, (b) $E_{PAD}$ \textit{vs} $T_r$, (c) $E_{PAD}$ \textit{vs} 80-30\% CFD. The cuts were rotated in order to project on the axis perpendicular to the identification curve. The values of $FoM$ for the three identification methods calculated for the energy ranges of 28-30, 33-35, and 38-40~MeV are presented in panel (d).}
\end{figure}
%------------------------------------   TAB. 2   ---------------------------------------------
\begin{table*}
\centering
\caption{\label{tab2} The obtained with three different methods values of $FoM$ for the $^{9,7}$Be isotopes are presented (columns 2-4) for the 2-MeV wide energy cuts centered at the energy values given in column 1.}
\begin{tabular}{llcc}
\hline
\hline
$E_{PAD}$ [MeV] & $FoM$ $I_{max}$ [a.u.]& $FoM$ $T_r$ [a.u.] & $FoM$ 80-30\% CFD [a.u.]\\
\hline									
29	& 0.61(14)	& 0.67(15)	& 0.98(8)\\
\hline									
34	& 1.23(9)	  & 0.93(10)	& 0.91(12)\\
\hline									
39	& 1.32(18)	& 1.13(11)	& 0.98(24)\\
\hline
\hline
\end{tabular}
\end{table*}
%---------------------------------------------------------------------------------------------

The correlation matrix between the energy and the $I_{max}$ constructed from the data recorded digitally by the first TRACE detector is presented in Fig.~\ref{matrices}(a). The energy scale is shown up to $\sim$76~MeV, while saturation was reached at around 93~MeV. The discrimination in $Z$ was easily achieved for protons, alpha particles, Li, Be, B, and C isotopes. The separation in $A$ is observed as well for the $^{7,6}$Li and, with minor statistics, for $^{9,7}$Be. The $^{8}$Be nucleus is unbound and therefore cannot be seen. The additional line in the Li region of the matrix, that reaches up to $\sim$1450~a.u. of $I_{max}$ and up to $\sim$38~MeV was interpreted as coming from the events when two alpha particles were detected at the same time in the same pad giving the summed signal.

The matrix of the energy $vs$ rise time (30-70\%) correlation, extracted from the signal of the same TRACE module, is shown in Fig.~\ref{matrices}(b). Event distributions for
alphas, $^{7,6}$Li and $^{9,7}$Be are again clearly separated, and events coming from two-alpha particles, simultaneously detected, are also observed on the left side of the Li distribution.

In both matrices of Fig.~\ref{matrices}(a) and (b) one can notice a significant background below 20~MeV, caused by alpha particles. Such background strongly limited the capability of the products separation at the lowest energies and will be discussed in the next section.

In Fig.~\ref{matrices}(c) the results of the analysis of the data from the second TRACE detector, connected to the analog acquisition chain, are presented. In this case, energies up to $\sim$44~MeV were recorded. The correlation matrix of the energy and the 80-30\% CFD allows to distinguish between the protons, alpha particles, $^{7,6}$Li, $^{9,7}$Be, B, and C reaction products. The double-alpha events are seen in vicinity of the Li curve, as previously. 

In order to quantify the mass separation capability using both digital and electronic chains, a figure of merit ($FoM$) was calculated for the Li and Be isotopes, for which sufficient statistics was collected. The $FoM$ value was defined as in Ref.~\cite{FoM}, i.e.:
$$ FoM = {{C_2 - C_1} \over {FWHM_1+FWHM_2}}~, $$
where $C_1$ and $C_2$ are the positions of centroids of the peaks corresponding to the $^{7}$Li and $^{6}$Li isotopes (Fig.~\ref{Fig2}) or $^{9}$Be and $^{7}$Be isotopes (Fig.~\ref{Fig3}). 

Projections of the 400-keV cuts centered at 20, 24.5, and 28.6~MeV for Li isotopes and the 2-MeV cuts centered at 29, 34, 39~MeV for Be isotopes, respectively, were done to obtain $FoM$ values from the identified peaks. In case of Be isotopes identification the cuts were projected on the axes perpendicular to the identification curves of interest, in order to reduce the energy spread caused by the wide energy windows used in the analysis. Using of such wide windows was necessary to improve the statistics.

The obtained values of $FoM$ for the $^{7,6}$Li isotopes are reported in Fig.~\ref{Fig2} and Tab.~\ref{tab1} for the different cuts and methods of identification. Figure~\ref{Fig2}(d) presents the dependence of the $FoM$ values on the energy at which the cut was cenetered. It is found that the value $FoM=0.75$, usually considered as allowing for a good separation of the peaks~\cite{FoM}, is achieved for the cut at 28.6~MeV in the $E_{PAD}$ \textit{vs} $I_{max}$ and $E_{PAD}$ \textit{vs} $T_r$ matrices (digital acquisition, $FoM=0.76$ for both methods) and for all the cuts starting from 20~MeV and above in the $E_{PAD}$ \textit{vs} 80-30\% CFD matrix (analog acquisition). However, one should note that there is an additional widening of the peaks caused by the large energy window taken to construct the spectra ($\Delta E_{PAD}=400$~keV). Therefore, the results obtained even for 20~MeV, i.e., $FoM$ values of 0.64(18) and 0.53(29) for $E_{PAD}$ \textit{vs} $I_{max}$ and $E_{PAD}$ \textit{vs} $T_r$, respectively, still allowing for identification of the $^{7,6}$Li isotopes, can be considered as approaching the lower limit of energy in terms of light products identification. We note that in the case of the $E_{PAD}$ \textit{vs} 80-30\% CFD method, applied for the signal from the second TRACE detector, the identification was achieved down to 19.8~MeV with $FoM=0.47(10)$.

The identification of $^{9,7}$Be isotopes in terms of extracted $FoM$ values was performed using 2-MeV wide cuts, to compensate the minor statistics. The cuts were centered at 39, 34, and 29~MeV. The extracted $FoM$ values for the three identification methods are reported in Fig.~\ref{Fig3}(a)-(c) and Tab.~\ref{tab2}, while the plot $FoM$($E_{PAD}$) is presented in Fig.~\ref{Fig3}(d). For the lowest cut at $E_{PAD}=29$~MeV, $FoM$ values of 0.61(14) and 0.67(15) for the $E_{PAD}$ \textit{vs} $I_{max}$ and $E_{PAD}$ \textit{vs} $T_r$ methods (digital acquisition) were obtained, respectively, showing a good identification ability, while the $FoM=0.98(8)$ value obtained for the analog $E_{PAD}$ \textit{vs} 80-30\% CFD method suggests that the lower limit of identification of Be isotopes has not been reached. However, projections at lower energies could not be used due to the very low statistics.

%---------------------------------------------------------------------------------------------
%----------------------------------   DISCUSSION   -------------------------------------------
%---------------------------------------------------------------------------------------------
\section{Discussion}
The limits for low-energy light ions identification by PSA technique obtained in the present experiment are below 20~MeV for Li and below 29~MeV for Be isotopes. These values could be further improved by reducing the background from alpha particles, which strongly limits the sensitivity below 20~MeV.

This background appears most probably due to the increased noise caused by radiation damage of the detector. We checked that in the matrices constructed from the data recorded during a short run at the beginning of the measurements this kind of background was not present. The question arose why in the analog processing of the signal this effect seems not to be so pronounced. We would interpret it as follows: the quantity used in the analog data analysis, that is the difference between 80\% and 30\% CFD, comes from the integration, thus, the noise is filtered. On the contrary, the method applied in PSA is by definition much more sensitive to the noise because it relies on the difference of the two interpolated values of the signal in case of $T_r$ or the derivative of the interpolated function ($I_{max}$). In this way, the larger noise at the end of the experiment must affect significantly the extracted $T_r$ and $I_{max}$. Moreover, the 100-MHz sampling rate giving the measured signal at every 10~ns limits the precision of extracting $I_{max}$ and $T_r$ which cannot be further improved by increasing the number of interpolated points.

The differences between the performances of the two acquisition systems used in the present experiment may come also from the fact that two separate signals from different detectors were processed. For example, the digitally read detector suffered from the leakage current of around 1.1~$\mu$A at the end of the experiment, while the smaller value, 0.9~$\mu$A, was observed for the second module. Moreover, we cannot exclude that the ideal symmetry in position of both detectors with rescpect to the beam has not been reached, which would result in slightly different level of radiation damage of the used modules.
%---------------------------------------------------------------------------------------------
%------------------------------------   SUMMARY   --------------------------------------------
%---------------------------------------------------------------------------------------------
\section{Summary}
We presented the results of the test experiment aiming at identification of the light products of the transfer reactions by using two 200-$\mu$m thick silicon TRACE detectors. Two acquisition systems, digital and analog, were employed to process the signals from the detectors. The digitized charge signal from one detector allowed to obtain the energy, maximum current ($I_{max}$), and rise time ($T_r$) values, while the processing of analog signal from the second detector provided energy and two time information signals from 30\% and 80\% Constant Fraction Discriminator (CFD). The correlation matrices of energy \textit{vs} $I_{max}$, $T_r$, and 80-30\% CFD confirmed the high quality of the Pulse Shape Analysis technique which can be used for the light ion identification as comparable with the analog approach.

The separation between the $^{7,6}$Li isotopes was obtained down to the energy of $\sim$20~MeV. Below that energy the significant background coming form alpha particles appeared, especially for the detector with the digital electronics, diffusing the extracted $I_{max}$ and $T_r$. The $^{9,7}$Be isotopes were separable down to the energy of $\sim$29~MeV. Therefore, the value of $\sim$3~MeV/nucleon can be considered as approaching the lower limit of energy in terms of light products identification by using single-layer silicon TRACE detectors.

%---------------------------------------------------------------------------------------------
%-------------------------------   ACKNOWLEDGEMENTS   ----------------------------------------
%---------------------------------------------------------------------------------------------
\section*{Acknowledgements}
This work was supported by the Polish National Science Centre under Contract No. 2014/14/M/ST2/00738 (COPIN-INFN Collaboration).

%---------------------------------------------------------------------------------------------
%----------------------------------   REFERENCES   -------------------------------------------
%---------------------------------------------------------------------------------------------

\end{document}